\PassOptionsToPackage{unicode}{hyperref}
\PassOptionsToPackage{hyphens}{url}
\documentclass[
  12pt,
]{article}
\usepackage{xcolor}
\usepackage[margin=1in]{geometry}
\usepackage{amsmath,amssymb}
\usepackage{iftex}
\ifPDFTeX
  \usepackage[T1]{fontenc}
  \usepackage[utf8]{inputenc}
  \usepackage{textcomp} 
\else 
  \usepackage{unicode-math} 
  \defaultfontfeatures{Scale=MatchLowercase}
  \defaultfontfeatures[\rmfamily]{Ligatures=TeX,Scale=1}
\fi
\usepackage{lmodern}
\ifPDFTeX\else
\fi
\IfFileExists{upquote.sty}{\usepackage{upquote}}{}
\IfFileExists{microtype.sty}{
  \usepackage[]{microtype}
  \UseMicrotypeSet[protrusion]{basicmath} 
}{}
\usepackage{graphicx}
\makeatletter
\newsavebox\pandoc@box
\newcommand*\pandocbounded[1]{
  \sbox\pandoc@box{#1}%
  \Gscale@div\@tempa{\textheight}{\dimexpr\ht\pandoc@box+\dp\pandoc@box\relax}%
  \Gscale@div\@tempb{\linewidth}{\wd\pandoc@box}%
  \ifdim\@tempb\p@<\@tempa\p@\let\@tempa\@tempb\fi
  \ifdim\@tempa\p@<\p@\scalebox{\@tempa}{\usebox\pandoc@box}%
  \else\usebox{\pandoc@box}%
  \fi%
}
\def\fps@figure{htbp}
\makeatother
\usepackage[utf8]{inputenc}
\usepackage{amsmath}
\usepackage{amsfonts}
\usepackage{amssymb}
\usepackage{booktabs}
\usepackage{graphicx}
\usepackage[hyphens]{url}
\usepackage{hyperref}
\hypersetup{colorlinks=true, linkcolor=black, citecolor=black, urlcolor=blue, pdfborder={0 0 0}}
\usepackage{libertinus}
\usepackage{setspace}
\usepackage[bottom,flushmargin,hang,multiple]{footmisc}
\usepackage{tabularx}
\usepackage{titlesec}
\usepackage{textcase}
\usepackage{threeparttable}
\usepackage{tocloft}
\usepackage[most]{tcolorbox}

\renewcommand{\thesection}{\Roman{section}} 
\renewcommand{\thesubsection}{\Alph{subsection}} 
\renewcommand{\thesubsubsection}{\arabic{subsubsection}} 
\renewcommand{\theparagraph}{\alph{paragraph}} 
\renewcommand{\thesubparagraph}{(\roman{subparagraph})} 
\titleformat{\section}{\normalfont\Large\scshape\centering}{\thesection.}{1em}{}
\titlespacing*{\section}{0pt}{3.5ex plus 1ex minus .2ex}{2.3ex plus .2ex}
\titleformat{\subsection}{\normalfont\large\bfseries}{\thesubsection.}{1em}{}
\titlespacing*{\subsection}{0pt}{3.25ex plus 1ex minus .2ex}{1.5ex plus .2ex}
\titleformat{\subsubsection}{\normalfont\normalsize\itshape}{\thesubsubsection.}{1em}{}
\titlespacing*{\subsubsection}{0pt}{3.25ex plus 1ex minus .2ex}{1.5ex plus .2ex}
\titleformat{\paragraph}[runin]{\normalfont\normalsize\itshape}{\theparagraph.}{0.5em}{}[\hspace{0.5em}]
\titlespacing*{\paragraph}{0pt}{1.5ex plus .5ex minus .2ex}{1em}
\titleformat{\subparagraph}[runin]{\normalfont\normalsize\itshape}{\thesubparagraph}{0.5em}{}[\hspace{0.5em}]
\titlespacing*{\subparagraph}{0pt}{1.5ex plus .5ex minus .2ex}{1em}
\usepackage{bookmark}
\IfFileExists{xurl.sty}{\usepackage{xurl}}{} 
\makeatletter
\@ifundefined{xmpquote}{}{}
\makeatother
\hypersetup{
  pdftitle={Private Again: Artificial Intelligence Agents Restore Anonymity---Foreclosing Discrimination and Its Proof},
  hidelinks,
  pdfcreator={LaTeX via pandoc}}

\title{Private Again: Artificial Intelligence Agents Restore
Anonymity---Foreclosing Discrimination and Its Proof}
\author{}
\date{\vspace{-2.5em}}

\begin{document}
\maketitle

\vspace*{\fill}

\begin{center}
Anirban Mukherjee

Hannah Hanwen Chang

\bigskip

Date: \today

Forthcoming, \textit{Stetson Law Review Forum} (Summer 2026)
\end{center}

\vspace*{\fill}

\noindent \hrulefill

\noindent Anirban Mukherjee (anirban@avyayamholdings.com) is Principal at Avyayam Holdings. Hannah H. Chang (hannahchang@smu.edu.sg; corresponding author) is Associate Professor of Marketing at the Lee Kong Chian School of Business, Singapore Management University. This research was supported by the Ministry of Education (MOE), Singapore, under its Academic Research Fund (AcRF) Tier 2 Grant, No. MOE-T2EP40124-0005.

\newpage

\begin{center}
\textbf{Abstract}
\end{center}
\singlespacing

\noindent Artificial intelligence agents can transact online on behalf
of a human principal---browsing, paying, receiving, and
reviewing---without revealing who that principal is. That architecture
starves algorithmic discrimination of its inputs---identity, purchase
history, location history, behavioral traces, and demographic
proxies---but also forecloses its proof. Disparate-treatment needs
comparators; disparate-impact needs protected-class baselines; and
\emph{Iqbal}-era pleading needs specific factual allegations---doctrinal
predicates that anonymous transactions never generate. The effects fall
asymmetrically: those most vulnerable to discrimination are least able
to afford the shield and, when harms remain, least able to prove them.
The challenge for the law shifts from detecting and remedying
algorithmic discrimination to governing agent-mediated anonymity as
civil rights infrastructure: ensuring access to privacy-preserving
agents, regulating abuse without forced identification, and deciding
whether retailers may refuse to deal with agents at all.

\begin{center}\rule{0.5\linewidth}{0.5pt}\end{center}


\smallskip


\newpage

\bigskip

\section*{Introduction}\label{introduction}
\addcontentsline{toc}{section}{Introduction}

\begin{quote}
``Nothing was your own except the few cubic centimeters inside your
skull.''

---George Orwell, \emph{Nineteen Eighty-Four}\footnote{\textsc{George Orwell},
  \textsc{Nineteen Eighty-Four} 28 (Harcourt, Brace \&
  Co.~1949).\label{fn:orwellfirst}}
\end{quote}

\bigskip

Each wave of technological change arrives with the same dread: it will
hollow out the private sphere.\footnote{\emph{See generally}
  \textsc{Daniel J. Solove}, \textsc{Understanding Privacy} 4--5 (2008)
  (tracing recurring privacy anxieties across successive information
  technologies).\label{fn:anxiety}} The printing press prompted fears
that ideas would be surveilled at the point of publication.\footnote{\emph{See}
  \textsc{Elizabeth L. Eisenstein},
  \textsc{The Printing Press as an Agent of Change} 346--48 (1979)
  (describing the Archbishop of Mainz's 1485 licensing edict and
  subsequent papal censorship decrees prompted by the printing
  press).\label{fn:printing}} The telegraph and the telephone put voices
into wires that, it was imagined, someone was always listening
to.\footnote{\emph{See} Olmstead v. United States, 277 U.S. 438, 473
  (1928) (Brandeis, J., dissenting) (warning that government wiretapping
  could obtain ``what is whispered in the closet'');
  \textsc{Tom Standage},
  \textsc{The Victorian Internet: The Remarkable Story of the Telegraph and the Nineteenth Century's On-line Pioneers}
  110--11 (1998) (describing widespread anxiety that telegraph operators
  and intermediaries could read every private message in
  transit).\label{fn:wiretap}} The industrial city was an alienating
nightmare to those who remembered the village.\footnote{\emph{See}
  \textsc{Raymond Williams}, \textsc{The Country and the City} 1--12
  (1973) (tracing the recurring literary contrast between an idealized
  rural past and a corrupt, alienating urban
  present).\label{fn:metropolis}} The credit card replaced the unmarked
handshake of cash with a permanent, centralized ledger of every
purchase.\footnote{\emph{See} \textsc{Josh Lauer},
  \textsc{Creditworthy: A History of Consumer Surveillance and Financial Identity in America}
  1--2, 184--86 (2017) (observing that each card payment enters ``an
  invisible surveillance network that confirms our identity {[}and{]}
  records the details of our transaction,'' and tracing the surveillance
  infrastructure the cashless society required).\label{fn:creditcard}}
Orwell's image of the last redoubt---those few cubic centimeters inside
the skull---is the most elegant expression of an anxiety that is, by
now, several centuries old.\footnote{Orwell, \emph{supra} note
  \ref{fn:orwellfirst}, at 28 (``Nothing was your own except the few
  cubic centimeters inside your skull.'').\label{fn:orwell}} In each
iteration, the fear is the same: the new machine sees too much,
remembers too well, and shares too widely.\footnote{\emph{See} Solove,
  \emph{supra} note \ref{fn:anxiety}, at 4--5.\label{fn:anxietyclose}}

Artificial intelligence (``AI''), however, holds the potential for an
inversion: a machine that lets its user be seen less, remembered less,
and traced less. As cryptocurrency has shown, the digital age need not
end anonymous exchange.\footnote{Bitcoin and its successors record
  transactions on a public ledger without necessarily linking them to
  real-world identities. \emph{See} Satoshi Nakamoto, \emph{Bitcoin: A
  Peer-to-Peer Electronic Cash System} 6 (unpublished manuscript),
  \url{https://bitcoin.org/bitcoin.pdf}
  {[}\url{https://perma.cc/QY6S-5EMX}{]} (describing a system where
  ``{[}t{]}he public can see that someone is sending an amount to
  someone else, but without information linking the transaction to
  anyone'') (last visited July 4, 2026);
  \textsc{Primavera De Filippi \& Aaron Wright},
  \textsc{Blockchain and the Law: The Rule of Code} 20--21
  (2018).\label{fn:crypto}} AI agents can extend this principle to a
commercial transaction as a whole: an AI agent can serve as an ephemeral
intermediary, indistinguishable from any other retail-platform user,
that browses, pays, receives, and reviews for a human principal without
revealing whom it represents.\footnote{\emph{See} Yonadav Shavit et al.,
  \emph{Practices for Governing Agentic AI Systems} 4 (2023),
  \url{https://cdn.openai.com/papers/practices-for-governing-agentic-ai-systems.pdf}
  {[}\url{https://perma.cc/58AG-D7GA}{]} (defining agentic AI systems as
  those that ``achieve complex goals . . . with limited direct
  supervision''); Noam Kolt, \emph{Governing AI Agents}, 101
  \textsc{Notre Dame L. Rev.} (forthcoming 2026) (manuscript at 8--10),
  \url{https://papers.ssrn.com/sol3/papers.cfm?abstract_id=4772956}
  {[}\url{https://perma.cc/5RL9-JVFL}{]} (analyzing AI agents through
  the lens of agency law because they autonomously act on behalf of
  human principals).\label{fn:agentdef}} It can pay in cryptocurrency or
with a single-use card, receive goods at an anonymized address, and
leave no readily accessible record linking the transaction to the
principal at the point of sale.\footnote{\emph{See supra} note
  \ref{fn:crypto}.\label{fn:agentanon}} Anonymity can be restored---not
by retreating from technology, but through it.

The civil rights consequences run in both directions. The architecture
of algorithmic discrimination---the profiles that tie each transaction
to an identifiable buyer and the models that infer protected traits from
the accumulated record\footnote{\emph{See infra} Part I.A (tracing the
  mechanisms and the data each consumes); on the composition and
  durability of the profiles, see \emph{infra} note \ref{fn:tracking};
  on the inference of protected traits from facially neutral data, see
  \emph{infra} notes \ref{fn:algorithmic}, \ref{fn:creditproxy}, and
  accompanying text.\label{fn:archsupport}}---collapses when the buyer
is a shell. A party that cannot identify its counterparty cannot
proxy-discriminate.\footnote{Proxy discrimination is the use of a
  facially neutral trait as a stand-in for a legally protected one.
  \emph{See} Anya E.R. Prince \& Daniel Schwarcz, \emph{Proxy
  Discrimination in the Age of Artificial Intelligence and Big Data},
  105 \textsc{Iowa L. Rev.} 1257, 1267 (2020) (``Proxy discrimination
  occurs when a facially-neutral trait is utilized as a stand-in---or
  proxy---for a prohibited trait.''). The practice was historically
  deliberate, a means of discriminating while evading antidiscrimination
  law, but it need not be. \emph{See id.} A model that mines neutral
  data for predictive power can settle on proxies for protected traits
  simply because they predict. \emph{See id.}\label{fn:proxydef}} Yet
that same collapse also defeats the evidentiary infrastructure on which
antidiscrimination enforcement depends. Disparate-treatment needs
comparators; disparate-impact needs protected-class baselines; and
\emph{Iqbal}-era pleading needs specific factual allegations.\footnote{Ashcroft
  v. Iqbal, 556 U.S. 662, 678 (2009).\label{fn:iqbalintro}} Anonymous
transactions do not generate these doctrinal predicates. The mechanism
that forecloses discrimination also forecloses its proof.

The challenge for the law shifts from detecting and remedying
algorithmic discrimination to governing the anonymity that forecloses
both. Part I of this Article traces how AI agents starve discrimination
of its inputs, proof of its evidence, and pleading of its facts. Part II
takes up three questions in turn---whether to treat agent-mediated
anonymity as civil rights infrastructure, how to regulate abuse without
forcing identification, and whether retailers may refuse to deal with
agents at all.

\section{Civil Rights Starved}\label{civil-rights-starved}

Civil rights law rests on an assumption so foundational it is rarely
stated: the parties to a transaction are identifiable, whether employer
and applicant, landlord and tenant, or---as here---seller and
buyer.\footnote{\emph{See} Samuel R. Bagenstos, \emph{The Structural
  Turn and the Limits of Antidiscrimination Law}, 94
  \textsc{Calif. L. Rev.} 1, 12 (2006) (analyzing antidiscrimination
  law's dependence on identifying perpetrators and victims of
  discriminatory conduct as a precondition for
  liability).\label{fn:assumption}} Discrimination, as the law
understands it, is something a seller does to a buyer that the seller
has categorized.\footnote{\emph{See} Int'l Bhd. of Teamsters v. United
  States, 431 U.S. 324, 335 n.15 (1977) (defining disparate treatment as
  when an employer ``simply treats some people less favorably than
  others because of their race, color, religion, sex, or national
  origin'' and noting ``{[}p{]}roof of discriminatory motive is
  critical'').\label{fn:teamsters}} Remedy is something a court imposes
after a plaintiff demonstrates that she was treated differently from a
similarly situated person outside her protected class.\footnote{McDonnell
  Douglas Corp.~v. Green, 411 U.S. 792, 807 (1973) (explaining that a
  court must order a prompt and appropriate remedy after finding
  unlawful discrimination).\label{fn:remedysupra}} Both presuppose that
the seller knows the buyer's identity and that the seller's knowledge
can be established.

This assumption is unremarkable in the traditional marketplace. Consider
a reader who wishes to buy a book that carries some social or
professional risk: a political memoir, a medical reference, or a work on
a stigmatized topic. In the ordinary online marketplace, the purchase
leaves a trail: the search query ties to an account, the order to a name
and billing address, the payment to a credit card, and the shipment to a
home.

Each trace is individually modest. But stitched together, they form a
durable profile---one that re-identifies the reader across sites and
visits. Subsequent browsing surfaces the same title or similar titles
across unrelated sites.\footnote{\emph{See} \textsc{Joseph Turow},
  \textsc{The Daily You: How the New Advertising Industry is Defining Your Identity and Your Worth}
  4--5 (2011). Several classes of identifier are used to track a user
  across sites: cookies and other stored identifiers, IP addresses, and
  device fingerprints derived from properties such as browser
  configuration, installed fonts, and screen resolution. The profile
  these identifiers assemble is durable in two senses: it persists,
  retained and traded by data brokers and ad networks across sessions
  and years; and it accretes, each transaction refining the record of
  identity, behavior, and inferred preference. \emph{See, e.g.,} Peter
  Eckersley, \emph{How Unique Is Your Web Browser?}, \emph{in}
  \textsc{Privacy Enhancing Technologies} 1, 2--5 (Mikhail J. Atallah \&
  Nicholas J. Hopper eds., 2010); Gunes Acar et al., \emph{The Web Never
  Forgets: Persistent Tracking Mechanisms in the Wild}, 2014
  \textsc{Procs. of the ACM SIGSAC Conf. on Comput. and Commc'ns Sec.}
  674, 675--76, 682--83; Steven Englehardt \& Arvind Narayanan,
  \emph{Online Tracking: A 1-Million-Site Measurement and Analysis},
  2016
  \textsc{Procs. of the ACM SIGSAC Conf. on Comput. and Commc'ns Sec.}
  1388.\label{fn:tracking}} Prices, goods, and services follow the
profile, as do advertisements. Different users with different profiles
receive different treatment.

An AI agent, however, alters the picture. It can present as an ordinary
user through an account the agent controls---rather than one tied to the
principal's identity---pay with a single-use virtual card or prefunded
cryptocurrency, and direct shipment to a parcel locker the principal
retrieves without showing identification linked to the
purchase.\footnote{\emph{See supra} note \ref{fn:agentdef}; \emph{infra}
  note \ref{fn:precursors}.\label{fn:mechsupra}} The retailer sees a
transaction indistinguishable from any other but cannot link that user
to the principal or to the principal's other transactions.\footnote{Each
  of the agent's steps has an analogue already in wide use:
  privacy-preserving browsing for tracking avoidance, virtual cards for
  payment, parcel lockers for receipt, and temporary email for
  communication. \emph{See, e.g.,} \textsc{Tor Project},
  \url{https://www.torproject.org}
  {[}\url{https://perma.cc/S9QJ-Y2J9}{]} (last visited July 4, 2026);
  \textsc{Privacy}, \url{https://privacy.com}
  {[}\url{https://perma.cc/ZML2-P94W}{]} (last visited July 4, 2026);
  \textsc{Amazon}, \url{https://www.amazon.com/b?node=6442600011}
  {[}\url{https://perma.cc/WU2H-A7ZQ}{]} (last visited July 4, 2026);
  \textsc{Proton}, \emph{SimpleLogin}, \url{https://simplelogin.io}
  {[}\url{https://perma.cc/QH3L-NCZF}{]} (last visited July 4,
  2026).\label{fn:precursors}}

The consequences are unusual precisely because they cut both ways. If a
retailer cannot identify the individual behind a transaction, it cannot
proxy-discriminate. It cannot price by zip code, steer by inferred
demographics, or target by protected class, for there is no data from
which to infer when the counterparty is an anonymous shell.\footnote{\emph{See
  infra} notes \ref{fn:zip}, \ref{fn:steering}, \ref{fn:creditproxy},
  and \ref{fn:adtargeting}.\label{fn:proxysupra}} The architecture of
algorithmic discrimination collapses. Anonymity, long thought a threat
to civil rights, becomes their champion.

Yet the same collapse dismantles civil rights law's remedial
architecture. Antidiscrimination enforcement depends on showing that a
protected class was treated differently, which presupposes identifying
who was treated, by whom, and how.\footnote{\emph{See, e.g.,}
  \emph{McDonnell Douglas Corp.}, 411 U.S. at 800--02 (setting forth the
  prima facie elements required to establish a Title VII discrimination
  claim).\label{fn:disptrtsupra}} Affirmative remedies depend on
identifying their beneficiaries. The mechanism that forecloses
discrimination forecloses its proof and its remedy.

\subsection{Discrimination Starved of Its
Inputs}\label{discrimination-starved-of-its-inputs}

The discrimination that civil rights law confronts in modern commerce is
principally statistical and algorithmic\footnote{\emph{See} Solon
  Barocas \& Andrew D. Selbst, \emph{Big Data's Disparate Impact}, 104
  \textsc{Calif. L. Rev.} 671, 673--77 (2016) (describing how
  contemporary discrimination arises from facially neutral data-mining
  practices); Pauline T. Kim, \emph{Data-Driven Discrimination at Work},
  58 \textsc{Wm. \& Mary L. Rev.} 857, 860--61, 890--92 (2017)
  (characterizing algorithmic ``classification bias'' as a distinct form
  of modern workplace discrimination).\label{fn:algorithmic}}: price
discrimination keyed to zip-code models,\footnote{Jennifer
  Valentino-DeVries, Jeremy Singer-Vine \& Ashkan Soltani,
  \emph{Websites Vary Prices, Deals Based on Users' Information},
  \textsc{Wall St. J.}, Dec.~24, 2012, at A1 (documenting Staples.com's
  practice of varying prices based on a user's inferred location and
  proximity to competitor stores); Ryan Calo, \emph{Digital Market
  Manipulation}, 82 \textsc{Geo. Wash. L. Rev.} 995, 1002--04 (2014)
  (theorizing personalized and geographically targeted pricing as a form
  of algorithmic consumer harm).\label{fn:zip}} product steering driven
by personalization signals,\footnote{\emph{See, e.g.,} Aniko Hannak et
  al., \emph{Measuring Price Discrimination and Steering on E-commerce
  Web Sites}, 2014 \textsc{Proc. of the ACM Internet Measurement Conf.}
  305, 308--11.\label{fn:steering}} credit scoring that relies on
proxies for race and age,\footnote{\emph{See, e.g.,} Robert Bartlett et
  al., \emph{Consumer-Lending Discrimination in the FinTech Era}, 143
  \textsc{J. Fin. Econ.} 30, 42--44 (2022) (finding that algorithmic
  lenders charge Latinx and Black borrowers higher interest rates than
  otherwise-similar White borrowers, even when no human loan officer is
  involved); Prince \& Schwarcz, \emph{supra} note \ref{fn:proxydef}, at
  1273--76 (theorizing how AI systems trained on facially neutral data
  systematically reconstruct protected characteristics through
  correlated proxies).\label{fn:creditproxy}} and ad targeting built on
behavioral aggregation.\footnote{\emph{See, e.g.,} Latanya Sweeney,
  \emph{Discrimination in Online Ad Delivery}, 56
  \textsc{Commc'ns. of the ACM} 44, 47--48 (2013); Muhammad Ali et al.,
  \emph{Discrimination Through Optimization: How Facebook's Ad Delivery
  Can Lead to Biased Outcomes}, 3
  \textsc{Procs. of the ACM on the Hum.-Comput. Interaction}, Nov.~2019,
  at 12--15; Charge of Discrimination at 3--6, Sec'y, U.S. Dep't of
  Hous. \& Urb. Dev. v. Facebook, Inc., FHEO No.~01-18-0323-8 (Mar.~28,
  2019), \url{https://archives.hud.gov/news/2019/HUD_v_Facebook.pdf}
  {[}\url{https://perma.cc/3MJC-MGJJ}{]}.\label{fn:adtargeting}} Each
mechanism consumes the same input---demographic or behavioral data tied
to an identifiable buyer---producing discrimination as a byproduct. A
retailer cannot steer products away from a protected class without
identifying which buyers belong to that class. A lender cannot price a
loan by zip code without geolocating the applicant. An ad network cannot
target by inferred race if the inference has no profile to land on.

When the counterparty is an AI agent, the inputs are gone. The agent
does not disclose the principal's location, inferred demographics,
purchase history, or protected class. Each session begins without
history; each session ends without persistence. The agent is a shell: no
stable identity to profile, no durable link to the human behind it.
Proxy discrimination, a dominant modern form, collapses.\footnote{\emph{See}
  Prince \& Schwarcz, \emph{supra} note \ref{fn:proxydef}, at 1275--78.
  Prince and Schwarcz show that depriving an algorithm of a protected
  characteristic does not prevent proxy discrimination, because a
  sophisticated model ``will identify and use any available data that
  even partially proxies for th{[}at{]} information''---geolocation,
  spending patterns, family medical history. \emph{Id.} at 1278;
  \emph{see id.} at 1276. That mechanism, however, presupposes the
  presence of proxies. Under agent-mediated anonymity, the agent's shell
  withholds not merely the protected characteristic but all possible
  proxies. With nothing available from which to reconstruct, the model
  can do no better than to treat every anonymous counterparty alike---it
  cannot discriminate. This protection is structural rather than
  behavioral: agent-mediated anonymity removes the capacity to
  proxy-discriminate rather than relying on the seller's forbearance.
  The point holds, though, only insofar as the agent is a genuine shell.
  Device fingerprints, payment metadata, a shipping address, or
  cross-session linkage might defeat the anonymity the shell provides,
  reintroducing the linked signals on which reconstruction
  depends.\label{fn:dominantform}} The retailer cannot engage in it. The
algorithm has no signal to fit. The discrimination machinery, deliberate
or inadvertent, is starved of its substrate.

\subsection{Proof Starved of Its
Evidence}\label{proof-starved-of-its-evidence}

The same mechanism that defeats discrimination defeats the enforcement
apparatus built to detect and punish it. The disparate-treatment
doctrine requires the plaintiff to show she was treated worse than a
similarly situated comparator outside her protected class.\footnote{\emph{See}
  McDonnell Douglas Corp.~v. Green, 411 U.S. 792, 802--04 (1973)
  (establishing the burden-shifting framework for disparate-treatment
  claims).\label{fn:disptrt}} That showing requires identifying herself
(the plaintiff), the defendant's knowledge (what the defendant could
have known about her), and a comparator (someone similarly situated but
outside the class who was treated differently). Each step depends on
records that AI agents do not produce. The plaintiff identifies herself
only by stepping out of the anonymity regime and disclosing the
information the agent was built to conceal; even then, the defendant may
have no record of who she was during the transaction. The defendant
lacks the requisite knowledge because the transactional interface
provides nothing to know. And the comparator is an agent,
indistinguishable from the plaintiff's own agent by design.

The disparate-impact doctrine fares no better. \emph{Griggs v. Duke
Power Co.} and its progeny require proof that a facially neutral
practice produces statistically significant differential effects on a
protected class.\footnote{Griggs v. Duke Power Co., 401 U.S. 424, 431
  (1971); \emph{see also} Tex. Dep't of Hous. \& Cmty. Affairs v.
  Inclusive Cmtys. Project, Inc., 576 U.S. 519, 540 (2015) (affirming
  disparate-impact liability under the Fair Housing
  Act).\label{fn:griggs}} That proof requires identifying the protected
class within the defendant's customer base and measuring outcomes
against that identification. If every customer is an agent, there is no
protected class to identify, no demographic baseline to compare against,
and no statistical machinery to deploy. The doctrine does not fail
because the practice is fair. It fails because the data to test fairness
does not exist. Discovery does not save the case: records the defendant
does not keep cannot be produced, and a retailer who sees only agents
has no customer-level demographics to disgorge. Expert testimony
encounters the same wall. An expert retained to show that a retailer's
checkout flow produces differential outcomes across protected classes
needs the ground truth of user identity; a retailer whose records
contain only agents has no such ground truth to provide. The expert is
left testifying about a pattern in data that, under an anonymity regime,
does not exist.

\subsection{Pleading Starved of Its
Facts}\label{pleading-starved-of-its-facts}

The problem with enforcement is not only that the evidentiary record is
thin at trial. It is that the record does not exist at the threshold at
which enforcement begins. \emph{Iqbal} and \emph{Twombly} elevated the
pleading standard from ``conceivable'' to ``plausible,'' requiring a
complaint to allege concrete facts from which unlawful conduct can be
reasonably inferred.\footnote{Ashcroft v. Iqbal, 556 U.S. 662, 678
  (2009); Bell Atl. Corp.~v. Twombly, 550 U.S. 544, 570
  (2007).\label{fn:iqbal}} A plaintiff must come to the courthouse with
facts sufficient to make discrimination a \emph{plausible}, not merely
conceivable, explanation for the defendant's conduct.\footnote{\emph{See}
  cases cited \emph{supra} note \ref{fn:iqbal}.\label{fn:plausiblesupra}}

Those facts come from three sources. The plaintiff may have \emph{direct
evidence}: a memo, an email, a policy, or a remark revealing
bias.\footnote{\emph{See} Price Waterhouse v. Hopkins, 490 U.S. 228,
  250--52 (1989) (plurality opinion) (recognizing that a plaintiff may
  prove discrimination by showing that an illegitimate criterion
  ``played a motivating part'' in the employment decision, including
  through direct expressions of bias); Michael J. Zimmer, \emph{The New
  Discrimination Law:} Price Waterhouse \emph{Is Dead, Whither}
  McDonnell Douglas\emph{?}, 53 \textsc{Emory L.J.} 1887, 1888--907
  (2005) (tracing the doctrinal role of direct evidence in
  disparate-treatment cases and the development of the ``stray remarks''
  doctrine in the lower courts).\label{fn:directevidence}} She may have
\emph{comparative evidence}: knowledge that similarly situated
comparators outside her protected class were treated
differently.\footnote{\emph{See} \emph{McDonnell Douglas}, 411 U.S. at
  804 (suggesting that a plaintiff may show pretext through evidence
  that ``white employees involved in acts against {[}the employer{]} of
  comparable seriousness . . . were nevertheless retained or rehired'');
  Suzanne B. Goldberg, \emph{Discrimination by Comparison}, 120
  \textsc{Yale L.J.} 728, 732--35 (2011) (analyzing the centrality and
  limits of the ``similarly situated'' comparator requirement in
  antidiscrimination doctrine).\label{fn:comparators}} Or she may have
\emph{statistical evidence}: patterns in a defendant's behavior revealed
by prior litigation, investigative journalism, regulatory disclosure, or
organized scrutiny.\footnote{\emph{See} Hazelwood Sch. Dist. v. United
  States, 433 U.S. 299, 307--08 (1977) (reaffirming that ``gross
  statistical disparities'' can themselves ``constitute prima facie
  proof of a pattern or practice of discrimination''); Int'l Bhd. of
  Teamsters v. United States, 431 U.S. 324, 339--40 (1977) (recognizing
  that ``statistics . . . come in infinite variety'' and may be
  probative of pattern-or-practice
  discrimination).\label{fn:statistical}} In a world of identifiable
customers, at least one of these is usually available. The architecture
of discrimination enforcement depends on that availability.

Agent-mediated anonymity removes all three sources at once. There is no
direct evidence because the agent strips the demographic signal that
would trigger discriminatory behavior: no email or memo can reveal bias
against a class the defendant never knew it was transacting with. There
is no comparative evidence because comparators are themselves anonymous
agents, indistinguishable from the plaintiff's own agent by
construction. And there is no statistical evidence because no defendant,
regulator, or third party can assemble demographic data about customers
who present only as shells. The pleading standard, faithfully applied,
results in dismissal, not because the plaintiff's claim is implausible,
but because the facts that would make \emph{any} such claim plausible
cannot exist.

The result is a doctrinal closed loop. A plaintiff cannot plead what she
cannot know; she cannot know what she cannot discover; she cannot
discover without first pleading. The gate locks from both sides.
\emph{Iqbal} is not malfunctioning; the Court's concern about abusive
pleading is met exactly as intended.\footnote{\emph{See} \emph{Iqbal},
  556 U.S. at 678--79.\label{fn:iqbalasintended}} But \emph{Iqbal}-era
pleading rigor combined with agent-mediated anonymity forecloses the
courthouse door to discrimination claims previously colorable under the
permissive pleading regime.\footnote{\emph{See} Conley v. Gibson, 355
  U.S. 41, 45--46 (1957) (establishing the ``no set of facts'' pleading
  standard subsequently abrogated by Bell Atl. Corp.~v. Twombly, 550
  U.S. 544, 570 (2007)).\label{fn:pleadingregime}} The enforcement
apparatus is not merely weakened. Its entry point is closed.\footnote{\label{fn:novelty}The
  argument that the pleading stage, rather than the ultimate burden of
  proof, is the operative failure point for discrimination claims in an
  anonymity regime appears novel to AI. Prior literature on
  discrimination pleading after \emph{Iqbal} has focused on the
  evidentiary difficulties faced by particular classes of plaintiffs in
  particular doctrinal settings, \emph{see, e.g.,} Joseph A. Seiner,
  \emph{The Trouble with Twombly: A Proposed Pleading Standard for
  Employment Discrimination Cases}, 2009 \textsc{U. Ill. L. Rev.} 1011,
  1014--15; Charles A. Sullivan, \emph{Plausibility Pleading Employment
  Discrimination}, 52 \textsc{Wm. \& Mary L. Rev.} 1613, 1621 (2011),
  without considering the combinatorial problem posed when the
  underlying identification architecture is absent altogether. The
  closed-loop structure described here parallels the ``intermediate
  copying double bind'' identified in the AI copyright context: claims
  cannot be initiated without the very evidence that the structure of
  the technology prevents from existing. \emph{See} Anirban Mukherjee \&
  Hannah Hanwen Chang, \emph{Pro Forma Copyright: AI Compaction and the
  Idea-Expression Impasse} 22 (May 6, 2026) (unpublished manuscript),
  \url{https://ssrn.com/abstract=6232359}
  {[}\url{https://perma.cc/4D9H-NTUP}{]}.}

The same closure affects the affirmative side of antidiscrimination law.
Antidiscrimination law is not only about preventing unequal treatment in
the moment. It also remedies the effects of historical marginalization
through targeted intervention, such as affirmative action in
contracting, minority business development programs, and targeted
lending initiatives.\footnote{\emph{See, e.g.,} 15 U.S.C. §§ 637(a),
  644(g)(1)(A)(iv); 12 U.S.C. § 2901.} Each presupposes that the
beneficiary can be identified as a member of the target group. At the
point of transaction, anonymity strips that identification: an agent
does not announce its principal's race to a minority business
development program; a cryptocurrency payment does not identify its
source community; a minority business set-aside cannot operate when
bidders are agents. At the transaction, the mechanism that prevents harm
also prevents help.

\section{The Role of Law}\label{the-role-of-law}

The return of anonymity to commerce poses three questions for the law:
whether anonymity should be treated as civil rights infrastructure,
available to those who cannot afford it; how to permit legitimate use
while foreclosing abuse; and whether retailers may refuse to transact
with agents at all---and if so, what that refusal means for the
anonymity this Article has described.

\subsection{Anonymity as Civil Rights
Infrastructure}\label{anonymity-as-civil-rights-infrastructure}

Sophisticated AI agents are not free.\footnote{\emph{See} Jiin Kim et
  al., \emph{The Cost of Dynamic Reasoning: Demystifying AI Agents and
  Test-Time Scaling from an AI Infrastructure Perspective},
  \textsc{arXiv}, Jan.~7, 2026, at 1--2, arXiv:2506.04301
  {[}\url{https://perma.cc/3B7E-8SSE}{]} (finding that agentic,
  multi-step AI workflows impose substantial resource, latency, energy,
  and infrastructure costs).\label{fn:agentcost}} They require technical
literacy, computational resources, and a willingness to navigate a
commercial world that has not yet adapted to them.\footnote{\emph{See}
  Lijia Ma et al., \emph{Learning to Adopt Generative AI},
  \textsc{arXiv}, Feb.~15, 2026, at 2--4, arXiv:2410.19806
  {[}\url{https://perma.cc/6QDA-W9MP}{]} (arguing that AI adoption and
  effective use vary according to education, technological background,
  and the ability to learn from experience).\label{fn:agentadoption}}
Early adopters will skew toward the populations that already enjoy the
best privacy tools: the wealthy, the technically fluent, and those with
the time to configure their transactional lives carefully.

This is the digital divide at a new layer. Agent-mediated commerce will
leave non-agent users exposed to the price discrimination, behavioral
targeting, and algorithmic harms that agent-users escape. The
discrimination machinery will not disappear; it will shift from
targeting members of a protected class to targeting the residual
population that did not, or could not, deploy the shield. The defeat of
discrimination against agent-users is not the defeat of discrimination;
it is its concentration on those who remain visible.

The obvious response is to sever the link between anonymity and wealth.
If privacy-enhancing agents become the most effective protection against
the discrimination machinery Part I described, access to those agents
should not be contingent on the ability to pay. Civil rights
infrastructure has addressed analogous problems before. Legal Aid exists
because legal representation is expensive; the Equal Employment
Opportunity Commission exists because individual enforcement is
resource-intensive; housing counselors exist because the information
asymmetries in mortgage and rental markets are severe.\footnote{\emph{See}
  \textsc{Alan Houseman \& Linda E. Perle},
  \textsc{Securing Equal Justice for All: A Brief History of Civil Legal Assistance in the United States}
  60 (Ctr. for L. \& Soc. Pol'y, rev. ed., 2018); 42 U.S.C. § 2000e-4
  (establishing the Equal Employment Opportunity Commission); 24 C.F.R.
  pt.~214 (2026) (HUD housing counseling program).\label{fn:lrsinfra}}
Each rests on the same principle: where a protection is necessary to
prevent a recognized form of discrimination, access to the protection
should not be limited to those who can afford it.

A parallel response would treat the AI agent itself as civil rights
infrastructure---publicly provided, subsidized, or required to be
offered free by platform providers above a certain scale. The analogues
are imperfect but not absent. Public libraries provide free internet
access because connectivity is now necessary for participation in civic
life. Universal service rules subsidize telecommunications in
underserved areas because the telephone network became too important to
be rationed by the ability to pay.\footnote{\emph{See} 47 U.S.C. § 254
  (universal service provisions added by the Telecommunications Act of
  1996); 47 C.F.R. pt.~54, subpt. E (2026) (Lifeline program subsidizing
  telecommunications service for low-income
  consumers).\label{fn:universalservice}}

Agent-mediated anonymity could be framed in the same register if the law
recognizes that what is being protected is not merely consumer
preference, but the structural precondition for civil rights enforcement
in a post-identification marketplace. Public libraries and universal
service are analogues for access to a positive good; the closest
precedent is \emph{Gideon v. Wainwright}'s right to counsel---a
protection against state power that the state itself
provides.\footnote{\emph{See} Gideon v. Wainwright, 372 U.S. 335, 344
  (1963) (``{[}R{]}eason and reflection require us to recognize that in
  our adversary system of criminal justice, any person haled into court,
  who is too poor to hire a lawyer, cannot be assured a fair trial
  unless counsel is provided for him.'').\label{fn:gideon}}

\subsection{Permitting Use, Foreclosing
Abuse}\label{permitting-use-foreclosing-abuse}

Anonymity has always been double-edged. Bitcoin enabled private
exchange, and it enabled the Silk Road marketplace and its
successors.\footnote{Silk Road was an online marketplace, operating on
  the Tor anonymity network and transacting exclusively in Bitcoin, that
  hosted anonymous sales of narcotics and other contraband until the
  government seized its servers and shut down the site in October 2013.
  \emph{See} United States v. Ulbricht, 858 F.3d 71, 82--83 (2d Cir.
  2017); \emph{see also} Nicolas Christin, \emph{Traveling the Silk
  Road: A Measurement Analysis of a Large Anonymous Online Marketplace},
  \emph{in} \textsc{Proc. of the 22d Int'l Conf. on World Wide Web} 213,
  220--22 (2013) (estimating Silk Road's annual revenue and transaction
  volume).\label{fn:silkroad}} End-to-end encryption protects the
dissident and the predator both.\footnote{\emph{See}
  \textsc{Steven Levy},
  \textsc{Crypto: How the Code Rebels Beat the Government---Saving Privacy in the Digital Age}
  197--98 (2001).\label{fn:cryptowars}} The history of privacy-enhancing
technology is, in significant part, a history of regulators negotiating
how much abuse to tolerate in exchange for how much legitimate
protection. Agent-mediated commerce will replay that negotiation at a
higher level of abstraction.

The distinctive difficulty is that agent-mediated anonymity is not a
shield around one dimension of the transaction---value, content,
identity---but around the transaction as a whole. A regulator responding
to Silk Road could focus on the payment rail: Bitcoin in, fiat out, and
know-your-customer (``KYC'') requirements at the on-ramps and off-ramps.
A regulator responding to encrypted messaging could focus on the
endpoint: the device, the app provider, and the metadata. A regulator
responding to agent-mediated commerce faces a technology whose function
is to make every point in the transaction opaque simultaneously. There
is no clean rail at which identification can be mandated. The shield has
no seams.

The KYC and anti-money-laundering (``AML'') regimes illustrate the
challenge. Existing financial regulation requires identification of
counterparties to certain transactions as a condition of participation
in the banking system.\footnote{Bank Secrecy Act, 31 U.S.C. §
  5318(\emph{l}) (requiring financial institutions to implement customer
  identification programs); 31 C.F.R. § 1020.220 (2026) (bank Customer
  Identification Program rule requiring identity-verification procedures
  sufficient to form a reasonable belief that the bank knows the
  customer's true identity); \textsc{Fin. Action Task Force},
  \emph{International Standards on Combating Money Laundering and the
  Financing of Terrorism \& Proliferation: The FATF Recommendations},
  Recommendation 10 (June 2026),
  \url{https://www.fatf-gafi.org/en/publications/Fatfrecommendations/Fatf-recommendations.html}
  {[}\url{https://perma.cc/46GS-NHHU}{]};
  \textsc{Fin. Crimes Enf't Network}, FIN-2019-G001, \emph{Application
  of FinCEN's Regulations to Certain Business Models Involving
  Convertible Virtual Currencies}, at 15--17 (May 9, 2019),
  \url{https://www.fincen.gov/sites/default/files/2019-05/FinCEN\%20Guidance\%20CVC\%20FINAL\%20508.pdf}
  {[}\url{https://perma.cc/ZHM6-RTQ5}{]}.\label{fn:kyc}} AI agents
transacting in cryptocurrency or single-use virtual card numbers sit
uneasily with these requirements. A response that requires the agent
itself to be identified defeats the architecture; one that exempts
agent-mediated transactions entirely would turn the agent into a
laundering tool.

By making every point in the transaction opaque simultaneously,
agent-mediated commerce compels the regulatory architecture to sit above
or below the transaction layer rather than within it.\footnote{Prior
  scholarship on cryptocurrency regulation, KYC/AML obligations, and
  platform governance has generally assumed that the transaction remains
  the operative regulatory unit. \emph{See, e.g.,} Sarah Jane Hughes \&
  Stephen T. Middlebrook, \emph{Advancing a Framework for Regulating
  Cryptocurrency Payments Intermediaries}, 32 \textsc{Yale J. on Regul.}
  495, 549--50 (2015); Kate Klonick, \emph{The New Governors: The
  People, Rules, and Processes Governing Online Speech}, 131
  \textsc{Harv. L. Rev.} 1598, 1635 (2018). Agent-mediated commerce,
  however, forecloses that assumption.\label{fn:regulatoryshift}} Two
directions are worth considering. Above the transaction, at the gateway,
agent providers might bear obligations to maintain audit trails
releasable only upon judicial authorization, thereby preserving a
pathway to identification when circumstances warrant. Below the
transaction, at the remedial layer, victims of agent-enabled fraud might
be made whole through ex ante insurance or distributive compensation
rather than through retrospective investigation, which anonymity renders
impractical.

\subsection{The Retailer Ban}\label{the-retailer-ban}

The private side of the market is likely to be less accommodating.
Retailers have strong incentives to identify their customers.
Identification enables personalized pricing, targeted upselling, profile
building, resale of behavioral data, and---not least---the commercial
surveillance and discrimination that Part I described as socially costly
but that are, from the retailer's perspective, profit-generating. A
retailer facing a customer who presents as an agent sees a customer
whose transactional value has been narrowed to the margin on the
immediate sale. Everything else the retailer would normally extract is
gone.

The natural response is refusal. A retailer can amend its Terms of
Service to ban agent-mediated transactions, require proof of human
identity before completing a purchase, implement CAPTCHA-like challenges
designed to defeat automation, or simply decline to serve any account it
suspects of being an agent. Each restores the retailer's
profile-building capacity by forcing the customer to present as a human
with a persistent identity. Under current law, each is presumptively
permissible: private actors generally set the terms of their own
commerce, and the idea that a retailer might be \emph{obligated} to
serve anonymous customers cuts sharply against the background
presumption of freedom of contract.\footnote{\emph{See generally}
  \textsc{P.S. Atiyah},
  \textsc{The Rise and Fall of Freedom of Contract} (1979) (tracing the
  rise and decline of contractual autonomy as an organizing principle of
  Anglo-American private law); \textsc{Lawrence M. Friedman},
  \textsc{Contract Law in America: A Social and Economic Case Study}
  (1965) (examining the social and economic forces shaping American
  contract doctrine).\label{fn:contractfreedom}}

But freedom of contract is not absolute. Public-accommodations law
already limits a retailer's freedom to refuse service on protected-class
grounds.\footnote{Civil Rights Act of 1964, tit. II, 42 U.S.C. § 2000a;
  \emph{see} Heart of Atlanta Motel, Inc.~v. United States, 379 U.S.
  241, 258--61 (1964); \emph{cf.} Masterpiece Cakeshop, Ltd.~v. Colo.
  C.R. Comm'n, 584 U.S. 617, 623--24 (2018) (addressing the tension
  between public-accommodations law and religious and free-speech
  objections to serving certain customers).\label{fn:title2}} If
agent-mediated anonymity is, as Part I argued, an effective protection
against algorithmic discrimination, then a retailer's refusal to serve
agents is a refusal to serve customers through a channel that prevents
discrimination. The functional consequence is the restoration of the
retailer's capacity to discriminate. The law's background
presumption---that private actors set their own terms---runs up against
the law's foreground commitment---that private actors may not set those
terms in ways that enable discrimination.

Framed narrowly, that is a business decision---a private actor setting
the terms of its own commerce. Framed more honestly, it is a demand for
identification as a condition of participation in the commercial sphere.
That demand has its own history in American civil rights law, and
history is not on its side. Pass laws in colonial and antebellum America
required free Black people to carry papers establishing their right to
be where they were; a person who could not produce them was presumed
enslaved and could be jailed, hired out, or sold.\footnote{\emph{See}
  \textsc{Ira Berlin},
  \textsc{Slaves Without Masters: The Free Negro in the Antebellum South}
  92--96, 327--40 (1974) (describing registration requirements,
  certificates of freedom, and pass laws imposed on free Black people
  throughout the antebellum South, and the jailing, hiring out, and sale
  of those found without papers); \textsc{Thomas D. Morris},
  \textsc{Southern Slavery and the Law, 1619--1860}, at 22--36 (1996)
  (tracing the legal definitions of race and the presumption that a
  Black person was enslaved, leaving freedom to be affirmatively
  proven).\label{fn:passlaws}} Voter identification debates continue to
turn on the disparate impact of identification requirements on the same
populations the civil rights regime is designed to protect; where courts
have found such requirements discriminatory, by result or by purpose,
they have held them unlawful.\footnote{\emph{See} Veasey v. Abbott, 830
  F.3d 216, 249--51 (5th Cir. 2016) (en banc) (holding that Texas's
  voter identification law had a discriminatory effect on minority
  voters in violation of Section 2 of the Voting Rights Act); N.C. State
  Conf. of the NAACP v. McCrory, 831 F.3d 204, 214--15 (4th Cir. 2016)
  (holding that North Carolina's photo-identification requirement and
  related voting restrictions were enacted with racially discriminatory
  intent in violation of Section 2 of the Voting Rights Act and the
  Fourteenth Amendment, and observing that the challenged provisions
  ``target African Americans with almost surgical precision''); Crawford
  v. Marion Cnty. Election Bd., 553 U.S. 181, 198--202 (2008) (plurality
  opinion) (upholding Indiana's voter-ID law but acknowledging the
  burden on voters who lack qualifying identification); Zoltan L.
  Hajnal, Nazita Lajevardi \& Lindsay Nielson, \emph{Voter
  Identification Laws and the Suppression of Minority Votes}, 79
  \textsc{J. Pol.} 363, 371--73 (2017). \emph{But see} Frank v. Walker,
  768 F.3d 744, 753 (7th Cir. 2014) (acknowledging findings that
  ``document a disparate outcome'' in identification possession but
  holding that Section 2 ``does not condemn a voting practice just
  because it has a disparate effect on minorities'').\label{fn:voterid}}
A commercial requirement that every purchaser be identifiable is
structurally analogous to these earlier regimes of forced
identification: imposed each time in the name of security, operating
each time as an instrument of exclusion.\footnote{The structural analogy
  between a retailer's Terms-of-Service ban on agent-mediated
  transactions and historical forced-identification regimes (pass laws,
  voter ID requirements) appears novel to AI commerce. Prior scholarship
  on algorithmic discrimination has generally focused on how systems use
  observed or inferred traits once individuals are legible to the
  system, \emph{see, e.g.,} Kim, \emph{supra} note \ref{fn:algorithmic},
  at 860--61; Anupam Chander, \emph{The Racist Algorithm?}, 115
  \textsc{Mich. L. Rev.} 1023, 1028--31 (2017), rather than on whether
  forced identification is itself a contested civil rights
  question.\label{fn:banasdiscrimination}}

\section{Conclusion}\label{conclusion}

Orwell's Winston Smith thought the inside of his skull was the last
uncolonized ground. This Article has argued that the perimeter of the
private can, in at least one dimension of modern life, be pushed back
outward again. AI agents can return to commerce a degree of anonymity
unavailable in practice since cash. The civil rights implications are
unusual: the mechanism of algorithmic discrimination collapses when the
counterparty is a shell, as does civil rights law's detection-and-remedy
architecture. Anonymity protects on one side and forecloses on the
other, falling hardest on the populations the civil rights project was
built to protect. Whether the law recognizes the possibility, and
whether it protects that possibility against private actors incentivized
to deny it, is the question next-generation commercial regulation must
face.

\end{document}